\documentclass[showpacs,superscriptaddress,preprintnumbers,amsmath,amssymb,aps,notitlepage,tight]{revtex4-1}
\usepackage[bookmarks=true,colorlinks=true,citecolor=blue,linkcolor=blue,urlcolor=magenta]{hyperref}
\usepackage{epsfig}
\usepackage{wrapfig}
\usepackage{amsmath}
\usepackage{dcolumn}% Align table columns on decimal point
\usepackage{multirow}
\usepackage{dcolumn}% Align table columns on decimal point
\usepackage{bm}% bold math
\usepackage{xfrac}
\usepackage{calc}
\usepackage{graphicx}
\usepackage[tight,footnotesize]{subfigure}
\usepackage{color}
\usepackage{longtable}%for tables that spans more than one page 
\usepackage{dashrule}%used for dashed lines in tables
\usepackage{multirow}%
\usepackage{rotating}%to turn words in the table cells
\usepackage{float}
\usepackage{wrapfig}
\usepackage{morefloats}%The package will al
\begin{document}
\preprint{AIP/123-QED}
\title{Smooth flux-sheets with topological winding modes}
\author{A. Bakry}
\affiliation{Institute of Modern Physics, Chinese Academy of Sciences, Gansu 730000, China}
%\author{X. Chen}%
%\affiliation{Institute of Modern Physics, Chinese Academy of Sciences, Gansu 730000, China}
\author{M. Deliyergiyev}\email[]{maksym.deliyergiyev@ujk.edu.pl}
\affiliation{Institute of Physics, The Jan Kochanowski University in Kielce, 25-406, Poland}
\affiliation{Institute of Modern Physics, Chinese Academy of Sciences, Gansu 730000, China}
\author{A. Galal}
\affiliation{Department of Physics,  Al Azhar University, Cairo 11651, Egypt\\Center of excellence for high energy physics, Adelaide, 5000 SA, Australia}
\author{M. KhalilA Williams}
%\affiliation{Department of Physics,  Al Azhar University, Cairo 11651, Egypt}
%\affiliation{Center of excellence for high energy physics, Adelaide, 5000 SA, Australia}
\affiliation{Department of Mathematics, Bergische Universit\"at Wuppertal, 42097 Germany}
\affiliation{Department of Physics, University of Ferrara, Ferrara 44121, Italy}
\affiliation{Research and computing center, The Cyprus Institute, Nicosia 2121, Cyprus}
%\author{ }
%\affiliation{Center for excellence for particle physics at tera scale, university of Adelaide, 5005 SA,Australia}
%\author{S. Xu}
%\affiliation{Institute of Modern Physics, Chinese Academy of Sciences, Gansu 730000, China}
%\author{P-M. Zhang}
%\affiliation{Institute of Modern Physics, Chinese Academy of Sciences, Gansu 730000, China}
\date{April 20, 2020}
\begin{abstract}  
  The inclusion of the Gaussian-curvature term in the bulk of Polyakov-Kleinert string action renders new boundary terms and conditions by Gauss-Bonnet theorem. Within a leading approximation, the eigenmodes of smooth worldsheets and the free-energy of a gas of open rigid strings appears to be altered at second order in the coupling by the topological term . In analogy to the topological $\theta$ term, the Gauss-Bonnet term is introduced into the effective action with a complex coupling to implement signed energy shifts. We investigate the rigid color flux-sheets between two static color sources near the critical point in the light of the topologically induced shifts. The Yang-Mills lattice data of the potential of static quark-antiquark $Q\bar{Q}$ in a heatbath is compared to the string potential. The Monte-Carlo data correspond to link-integrated Polyakov-loop correlators averaged over SU(3) gauge configurations at $\beta=6.0$. Substantial improvement in the fit behavior is displayed over the nonperturbative source separation distance $0.2$ fm to $1.0$ fm. Remarkably, the returned coupling parameter of the topological term from the fit exhibits a proportionality to a quantum number. These findings suggest that the manifested modes are the winding number of a topological particle on the string's worldsheet.
\end{abstract}
%PACS, the Physics and Astronomy %Classification Scheme.
%Use showkeys class option if keyword%display desired
\maketitle
% body of paper here
%%%%%%%%%%%%%%%
%\section{Introduction\label{sec:intro}}
\paragraph{{\bf \large{Introduction:}}}
%%%%%%%%%%%%%%%
   The relativistic string theory is a theory valid at all distance scales but well known of the short comings of the critical dimensions and tachyonic spectrum~\cite{NG}. Nevertheless, perturbative stability preserving the Lorentz symmetry can be established~\cite{polchinski-strominger} in the long string limit at a given order with the action expanded around the classical solution. 

  In pertinence to hadronic physics, the energy of the ground-state of the confining glue remains away from being string-like at $Q\bar{Q}$ source separations up to 0.5 fm~\cite{Luscher:2002qv}. It is presumably the distances over which the finite thickness  of the binding string brings forth a resistance to the transverse bending or a rigid aspect.   

  The rigidity conjecture had been ingrediented by Polyakov and Kleinert~\cite{PK}(PK) into the bosonic string theory as a possible model of the QCD flux-tube. In correspondence with the classical modes of thick strings, the smooth flux-sheets configurations of the PK model are more probable than those that the sharply creased by virtue of the string stiffness. This is contained by means of the parameters of the extrinsic curvature/shape tensor of the worldsheet surfaces of the string. 

  The extrinsic curvature term and the Gaussian-curvature are among the second derivative geometric-invariants allowed by Poincare symmetry and are naturally encompassed into the theory of L\"uscher and Weisz~\cite{LUSCH81}. The string model while coupled to extrinsic curvature can be tachyonic free above some critical coupling~\cite{Stability, ClosedSmooth}. Moreover, the model accommodates the main features of QCD such as the asymptotic freedom, infrared (IR) confinement~\cite{PK}, real $Q\bar{Q}$ potential~\cite{GK, Ambjorn:2014rwa} and the thermal deconfinement transition~\cite{TC}. 

  The Gaussian-curvature term in the action, on the other hand, recasts the boundary conditions of rigid string wave equation. The coupling parameter finely tunes the model of hadronic forces~\cite{Gauss-Bonnet, Acta} over relatively short but still within the non-perturbative distance scales. The intrinsic-curvature term endows the rigid string model some other favorable theoretical aspects~\cite{c4} such as removing the tachyonic state~\cite{c4} for configurations of massive ends. In the case of open strings the term inhibits the light-cone trajectories of the string at end points~\cite{c4}.
  
    With the unceasing refinement in the resolution of LGT numerical simulations, a reviving interest in divulging smooth strings in the hadron has been proclaimed in the recent literature~\cite{Ambjorn:2014rwa,Caselle:2014eka,Brandt:2017yzw,Bakry:2020flt}. The confining potential of the compact U(1) gauge group ~\cite{Caselle:2014eka} and non-abelian SU(N) gauge theories in 3D ~\cite{Brandt:2017yzw} is discussed. Though not dominant the rigid-like structure of the string condensate is numerically identified~\cite{Caselle:2014eka,Brandt:2017yzw,Bakry:2020flt} in the analysis of the static potential data, in addition to the energy profile in the vicinity of the critical point of the gluonic QCD~\cite{Bakry:2020flt}.

    The Gaussian curvature-term remained, nevertheless, obscure and sparsely envisaged from the numerical point of view. The intrinsic curvature is topological in 2D and its perturbative expansion has a leading term proportional to the EOM of Nambu-Goto string. These considerations deemed dropping the topological term to well suit long strings. Even so, the consideration of the leading order extrinsic-curvature term rigorously transforms the EOM~\cite{GK} from Laplace PDE into Helmholz PDE. The Gaussian curvature translates new boundary conditions imposed on the EOM of the (PK) rigid string and the gluon propagator~\cite{GK}. 
   
    The total surface integral of the intrinsic curvature in the action induces topological fluctuations~\cite{topologyterm} with subsequent modifications of the boundary conditions~\cite{Gauss-Bonnet} by virtue of the Gauss-Bonnet theorem. The coupling of the PK string action to the boundary curvatures~\cite{Gauss-Bonnet} could displace the energy spectrum to detectable levels over hadronic diameters $R < 0.5$ fm. This is typically the distance scale where the ``stiffness/resistance to bending/intrinsic-thickness'' of the flux tube is expected to be of considerable relevance to the model.
    
    In the following, we overview the defining equations of the string wave function and the free-energy of open string gas. We draw a comparison between numerical data of the static $Q\bar{Q}$ force in 4D  with the string potential of PK string with Gaussian-curvature at high temperature.
\paragraph{{\bf \large{The static potential:}}}
  The relativistic L\"uscher and Weisz (LW) action ~\cite{Caselle:2014eka} reads
\begin{equation}
  S^{\rm{LW}}=\sigma_{0} A+\dfrac{\sigma_{0}}{2} \int d^2\zeta \Bigg[\left(\dfrac{\partial \bm{X}}{\partial \zeta_{\alpha}} \cdot \dfrac{\partial \bm{X}}{\partial \zeta_{\alpha}}\right)+ \kappa_2 \left(\dfrac{\partial \bm{X}}{\partial \zeta_{\alpha}} \cdot \dfrac{\partial \bm{X}}{\partial \zeta_{\alpha}}\right)^2 +\kappa_3 \left(\dfrac{\partial \bm{X}}{\partial \zeta_{\alpha}} \cdot \dfrac{\partial \bm{X}}{\partial \zeta_{\beta}}\right)^2+... +\Bigg]+\int d^2\zeta\Bigg(\alpha \sqrt{-g} K^2+ i\theta  \sqrt{-g} R+..\Bigg).
\label{LW} 
\end{equation}
     where $\sigma_{0}$ and $\alpha$ are the string tension and rigidity parameter. The Gaussian-curvature coupling is $\gamma=i\theta$ where $\theta$ is complex in general. The term $S_{cl}$ characterizes the classical action. The vector $X_{\mu}(\zeta_{1},\zeta_{2})$ maps the area $ \mathcal{A} \subset \mathcal{\mathbb{R}}^{2} $ into $\mathcal{\mathbb{R}}^{4}$. The extrinsic and inner curvatures are defined as 
\begin{equation}
K= \dfrac{1}{\sqrt{g}}\partial_\alpha [\sqrt{g} g^{\alpha\beta}\partial_\beta] X,
\end{equation}
and
\begin{equation}
R= \left[g^{\alpha \beta} g^{\gamma \eta}-g^{\alpha \eta} g^{\beta \gamma}\right] \nabla_{\alpha} \nabla_{\beta} X_{\mu} (\nabla_{\gamma})^2 X^{\mu},
\end{equation}  
respectively. The two geometrical terms satisfy the Poincare and parity invariance and lies within~\cite{Caselle:2014eka} the general class of (LW) string actions~\eqref{LW} in the physical-gauge.

 With the approximations $\sqrt{-g} \simeq 1- \frac{1}{2} {\dot{\bf X}}^{2}+\frac{1}{2} {\bf X}^{2}, \text{and}\;\;\dfrac{1}{\sqrt{-g}}\simeq 1+\frac{1}{2}{\dot{\bf X}}^2-\frac{1}{2} {\bf X'}^2$, the Lagrangian density of the system assumes the form,
%\end{equation}
% = \sqrt{1-{\dot{\bf X}}^{2}+{\bf X'}^{2}} 
\begin{equation}
  \mathcal{L}=-\frac{1}{2} M_o^2 \left(\bm{X}^2-\dot{\bm{X}}^2\right)-\frac{1}{2} \alpha  \left(\bm{X}''-\ddot{\bm{X}}\right)^2-\gamma \ddot{\bm{X}} \cdot \dot{\bm{X}}-M_o^2. 
\label{LagDen}
\end{equation}
In the above the notation
% $\dot{\bf X}$ \A8 $\acute{\bf X}$ as small quantities
%\begin{equation}
$\dfrac{\partial {\bf X}}{\partial \zeta_0}=\dot{\bf X};~~\dfrac{\partial {\bf X}}{\partial \zeta_1}= {\bf X'}$ is used. The boundary conditions of the corresponding Euler-Largrange system ~\cite{IRST},
%\begin{equation}
%\frac{\partial  \mathcal{L}}{\partial  X_{\mu}}-\frac{\partial  }{\partial  \zeta_1} \left(\frac{\partial  \mathcal{L}}{\partial  X'_{\mu}}\right)-\frac{\partial  }{\partial  \zeta_0}\left(\frac{\partial  \mathcal{L}}{\partial  \dot{X}}\right)+\frac{\partial^2 }{\partial  \zeta_1^2}\left(\frac{\partial  \mathcal{L}}{\partial  X_{\mu}''}\right)+\frac{\partial ^2}{\partial  \zeta_0^2}\left(\frac{\partial  \mathcal{L}}{\partial  \ddot{X_{\mu}}}\right)=0,
%\end{equation}
\begin{equation}
\frac{\partial  \mathcal{L}}{\partial  X'_{\mu}}+\frac{\partial  }{\partial  X'_{\mu}}\frac{\partial  \mathcal{L}}{\partial  \zeta_1}=0;\quad \frac{\partial  \mathcal{L}}{\partial  X''_{i}}+\frac{\partial  }{\partial  X''_{\mu}}\frac{\partial  \mathcal{L}}{\partial  \zeta_1}=0.
\end{equation}
The Euler-Lagrange equation gives rise to the linear equations of motion
\begin{equation}
  M_o^2 \left(X''_{i}-\ddot{X} \right)-\alpha  \left(\dfrac{\partial^2 \ddot{X_i}}{\partial \zeta_{0}^{2}}-2 \dfrac{\partial^4 X_i}{\partial \zeta_{0}^{2}\partial \zeta_{1}^{2}}+ \dfrac{\partial^2 X''_{i}}{\partial \zeta_{1}^{2}}\right)=0.
\label{EOM}
\end{equation}
The boundary conditions are explicitly 
\begin{equation}
  \frac{\partial  }{\partial  \zeta_1}\left(X \left(-M_o^2\right)-\alpha  \left(X''-\ddot{X}\right)-\gamma  \ddot{X}\right)=0;\quad\quad \zeta_1=0 , \zeta_1=R;
\label{BC1}  
\end{equation}
and
\begin{equation}
  -\alpha  \left(X''-\ddot{X}\right)+\gamma  \ddot{X}=0;\quad\quad \zeta_1=0 , \zeta_1=R.
\label{BC2}  
\end{equation}
  The general solution to the string EOM, Eq.\eqref{EOM}, with characteristic roots are given as
\begin{equation}
 %\begin{split}
\Phi_{\mu}(\zeta_0,\zeta_1)= \sum\limits_{j=1}^{4}  A_{\mu}^{j}\; e^{i\omega\zeta_{0}+i k_{j} \zeta_1},\quad \quad  k_1=\omega,\;\;k_2=-\omega, \;\; k_3=\nu, \;\; k_4=-\nu,\;\;\text{and}\;\; \nu=\sqrt{\omega^2-M_0^2/\alpha},
\label{GS}
 % \end{split}
\end{equation}
respectively. The coefficients $A_{\mu}^{j}$ define the amplitudes. A system of linear equations is obtained by plugging the general solution Eq.~\eqref{GS} into the EOM, Eq.~\eqref{EOM} and using BCs of Eq.~\eqref{BC1} and Eq.~\eqref{BC2}. A solution is ensured provided the determinant vanishes at the roots of
\begin{equation}
u(\omega)=\sin(\nu R)\sin(\omega R)[(M_0^2-\gamma \omega^2)^4+\gamma^4\omega^6\nu^2]-2 (M_0^2-\gamma\omega^2)^2\gamma^2\omega^3\nu[1-\cos(\nu R)\cos(\omega R)].
\label{EigenF}
\end{equation}
The zeros of $u(z)$ are in general complex $\{\omega_n\} \subset \mathcal{C}$ and define eigenfrequencies of the string's wave function.

  The normalized free energy, $ \Lambda \rightarrow \infty$, is obtained in ~\cite{Just} and turns out to be 
\begin{equation}
F(R,T)= (d-2)T \sum _{n=-\infty }^{\infty }  \log \Bigg[u(i \,\omega_{n})] \Bigg]^{R}_{R=\infty}.
\end{equation}  
  The variation in the static potential owing to Gaussian curvature at one-loop order~\cite{Gauss-Bonnet} can be isolated as 
\begin{equation}
  \delta V(R,T)= \frac{4 \pi (d-2)\theta^2 T}{M^2} \sum_{n=0}^{\infty} \omega_n^3  \Omega_n \dfrac{\left(e^{-R \omega_n}-e^{-R \Omega_n}\right)^2}{\left(1-e^{-2R \bar{\omega}_n}\right) \left(1-e^{-2R\Omega_n}\right)},
\label{correction}  
\end{equation}
 with $\omega_{n}= 2n\pi T$ and $\Omega_n=\sqrt{(2n\pi T)^{2}+M^{2}}$. In correspondence to the topological $\theta$ term~\cite{Polyakov:1996nc, Unsal:2012zj}, the Gauss-Bonnet term is introduced into the effective action with a complex coupling to enable signed energy shifts Eq.~\eqref{correction}.
\begin{wrapfigure}{r}{0.3\textwidth}
\begin{flushleft}  
  \includegraphics[scale=0.24]{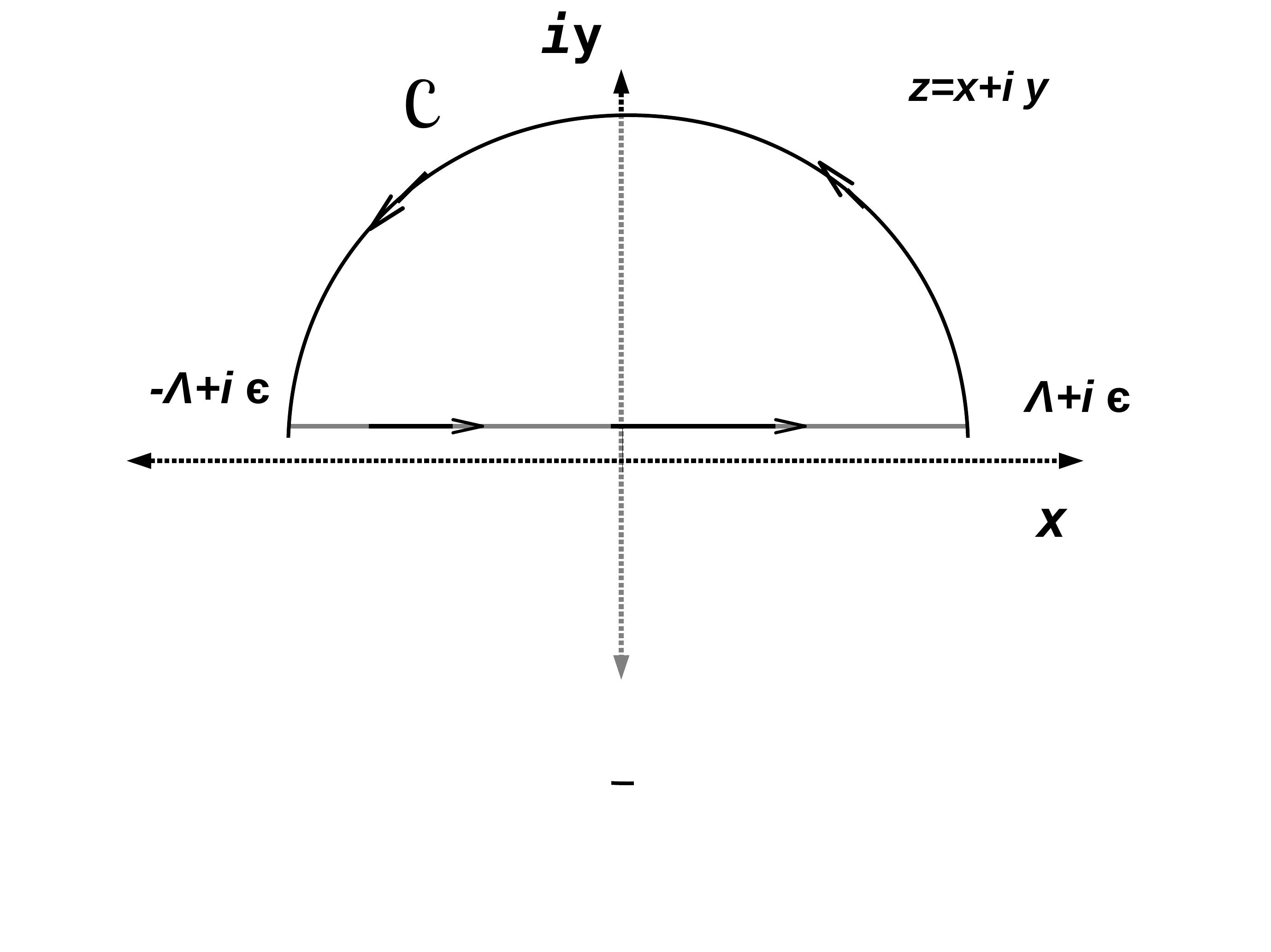}
%\vspace{-2.0 cm}    
\caption{\label{Fig0}The contour integral around eigenfrequencies.}
\end{flushleft}
%\vspace{-3.0 cm}  
\end{wrapfigure}
%\vspace{-3.1 cm}

  In addition to the leading-order correction Eq~\eqref{correction} a handy expression for the static potential setting $\gamma=0$ can be obtained from the partition function of the rigid string 
\begin{equation}
Z = \int D \bm{X} \exp -\sigma[ 1+\frac{1}{2} \bm{X}(1-\frac{\triangle}{2 M^2})(-\triangle)\bm{X}].
\label{2}
\end{equation}
Implementing the transformation $\bm{X}'=\dfrac{\triangle}{2 M^2}\bm{X}$, the partition function decouples to
\begin{equation}
Z^{(NG)}_{\ell o} Z^{(R)} =e^{-\sigma R T-\mu(T)} [\rm{Det}\left(-\triangle\right)]^{-\frac{(d-2)}{2}}[\rm{Det}\left(1-\frac{\triangle}{M^{2}}  \right)]^{-\frac{(d-2)}{2}},
\label{4}
\end{equation}
  where $M^{2}=\dfrac{\sigma_{0}}{2 \alpha}$ and $T=\dfrac{1}{L_{T}}$ is the temperature scale corresponding to the inverse of the finite extend in time direction. The eigenvalues and the functional trace of the two operators are given by
 \begin{equation}
%   \psi_{mn}=e^{2\pi i\left(\frac{m}{R}+\frac{n}{L_{T}} \right)},
   \lambda_{nm}=\left(\frac{2\pi n}{ L_{T}}\right)^{2}+\left(\frac{\pi m}{R}\right)^{2}, \quad \quad \xi_{nm}=\left(\frac{2\pi n}{ L_{T}}\right)^{2}+\left(\frac{\pi m}{R}\right)^{2}+M^{2},
\label{EigenV}
\end{equation}
respectively. Using the EOM and Eq.~\eqref{EigenV} the second trace is
\begin{equation}
%\begin{split}  
\log \left(\rm{Det}\left(1-\frac{\triangle}{M^{2}}  \right)\right)=  -\lim_{s \rightarrow 0}\frac{d}{ds}\sum_{n,m} \left(\frac{4\pi^{2}}{M^{2}} \left[\frac{m^{2}}{R^2}+\frac{n^2}{L_{T}^{2}} + \frac{M^{2}}{4 \pi^{2}} \right]\right)^s,
\label{9}
%\end{split}
\end{equation}
 where $s$ is an auxiliary parameter. The summation over $n$ in the above expression can be presented as a contour integral in the complex plane $z$ using Sommerfeld-Watson transform~\cite{ClosedSmooth} as
\begin{equation}
%\begin{split}  
\log(Z^{R})=\frac{(d-2)}{2} \Bigg(\lim_{s \rightarrow 0}\frac{d}{ds}\sum_{m} \Bigg[ \left(\frac{4\pi^{2}}{M^{2}R^2}\right)^{-s}\Bigg( \oint_{ c}dz \left(\dfrac{\exp(i\pi z)}{2i\sin(\pi z)}+\dfrac{1}{2} \right) \left(m^2+z^2+\frac{M^{2}R^2}{4\pi^{2}} \right)^{-s} \Bigg]\Bigg).
\label{10}
%\end{split}
\end{equation}
  The relevant poles to Eq.~\eqref{10} are located in the upper half-plane within the contour $C$ (Fig.~\ref{Fig0}).
\begin{equation}
%  \begin{split}
 \frac{2}{2-d} \log(Z^{R})=4 T \sum_{m=0}^{\infty} \log \left(1-e^{2\pi \tau\sqrt{n^2+\frac{M^2R^2}{4 \pi^2}}}  \right)  -\lim_{s \rightarrow 0}\frac{d}{ds}\left( \frac{4\pi^{2}}{M^{2}\,R^{2}} \right)^{-s} \frac{\sin(\pi s)}{\cos(\pi s)}(2\tau)^{1-2s}\frac{\Gamma^{2}(1-s)}{\Gamma(2-2s)} \sum_{m} \frac{1}{(m^2+M^{2}/4\pi^2)^{s-1/2}}.
 \label{11}
%\end{split}
\end{equation}
We proceed in the regularization of the second Epstein-Hurwitz $\zeta$ function using an integral representation of $\Gamma$ function as detailed in~\cite{Just,Bakry:2020flt}. An expression of the partition function would read
%, involving a series of Bessel functions, would read 
%, which is summed over $m$, the resultant expression for
%with the integral representation~\cite{Just} Bessel functions of the second kind $\sum_{n}n^{s-1/2}K_{s-1/2}(2 n M\,R)$ such that 
%\begin{equation}
%\begin{split}
%\zeta(s,M/4\pi)&=\frac{1}{\Gamma(s)}\int_{0}^{\infty} t^{s-1}\sum_{n=1}^{\infty} e^{-t(n^2+M^{2}\,R^2)}dt\\
%  &=\frac{-(M^2 R^2)^{-2s}}{2}+\frac{\sqrt{\pi} \Gamma(s-\frac{1}{2})}{2\Gamma(s)}(M^2 R^2)^{-s+\frac{1}{2}}+\frac{\sqrt{\pi}}{\Gamma(s)}\sum_{n=1}^{\infty} \int_{0}^{\infty} t^{s-\frac{3}{2}} \exp\left(-t \frac{M^{2}R^{2}}{\pi^{2}}-\frac{\pi^{2}n^{2}}{t}\right)dt
%  \label{12}
%\end{split}  
%\end{equation}
%  The integral in the last term can be presented in terms of sum over the modified Bessel functions of the second kind $\sum_{n}n^{s-1/2}K_{s-1/2}(2 n M\,R)$. The resultant expression for the partition function would read
\begin{equation}
%\begin{split}  
Z^{R} =\exp \Bigg[ \frac{(d-2)M}{2\pi}\sum_{n=1}n^{-1}K_{1}( 2 n M\,R) \Bigg] \prod^{\infty}_{n=0}\left(1-e^{2\pi \tau\sqrt{n^2+\frac{M^2R^2}{4 \pi^2}}}  \right).
%\end{split}
\label{14}
\end{equation}
 The potential corresponding to the total partition function $Z^{NG}_{\ell o} Z^{R}_{\ell o}$ is thus
\begin{equation}
%\begin{split}
V^{R}(R,T)=\sigma_{0} R+(d-2)T\Bigg[ \log\left(\eta(\tau)\right)+\frac{M}{2\pi T}\sum_{n=1}^{\infty}n^{-1}K_{1}( 2 n M\,R)]+\sum_{n=0}^{\infty} \log(1-e^{2\pi \tau\sqrt{n^2+\frac{M^2R^2}{4 \pi^2}}})+ \delta V(R,T,\alpha,\gamma)\Bigg]+\mu(T).
%\sum^{\infty}_{n=0} \Bigg[ \log \left(u(2n \pi T)   \right)\bigg]. 
%\end{split}
\label{20}
\end{equation}
$\mu$ is the ultraviolet (UV) cutoff scale and $\delta V(R,T)$ is given in accord to Eq.~\eqref{correction}. 
\paragraph{{\bf \large{Numerical discussion:}}}
\begin{wraptable}{r}{0.49\textwidth}
\caption{The $\chi^{2}$ values and fit parameters returned from fits to the string potential $V^{R}$ given by Eq.~\eqref{20}.}  
%  \begin{flushleft} 
%    \begin{minipage}[t]{0.4\textwidth}\flushleft	
		\begin{ruledtabular}
			\begin{tabular}{cc|cccccccccc}
%			  \multirow{2}{*}{\begin{turn}{90}\hspace{1cm}$ V_{{n\ell o}}^{R,b_2} $ \end{turn}} &\tiny{Fit Interval}
				&\multicolumn{2}{c}{\tiny{Fit Parameters, $T/T_{c}=0.9$}} &\multicolumn{2}{c}{}\\
				%&\multirow{2}{*}{$\chi^{2}$} &\multirow{2}{*} {$\mu$}
				%&\multirow{2}{*}{$\chi^{2}$} &\multirow{2}{*} {$\mu$}\\
				&$R \in I$
				&\multicolumn{1}{c}{$\chi^2$} &\multicolumn{1}{c}{$\sigma_{0}$ ($fm^{-2}$)} 
			        &\multicolumn{1}{c}{$ \alpha/a^{2}$} &\multicolumn{1}{c}{$\theta^2$}&\multicolumn{1}{c}{$\mu$}\\
				\hline
%\multirow{10}{*}{\begin{turn}{90} $V_{LO}$ \end{turn}}				
%\multirow{10}{*}{\begin{turn}{90}\hspace{1cm}$V_{{n\ell o}}+V^{R}+V^{b_2}$ \end{turn}}
%\multirow{10}{*}{\begin{turn}{90}\hspace{1cm}$V_{{n\ell o}}+V^{R}+V^{b_2}$ \end{turn}}				
% &0.045 &1083.06 &-0.446299\\ 				
%\multicolumn{1}{c}{} &0.038 & 1099.03   &      4.73    &   96.54      &  58.34        \\
 &\tiny{$[L_{m},L_{M}]$}  &    &            &     \\
%\multicolumn{1}{c}{} &[7,10]  & 19.42   & 8.99  &  0.038 &-0.37  &23.01     \\
\multicolumn{1}{c}{}&[2-6] &176.1   & 4.185(4)  &5.22(3)   &0          &-0.394(4)\\                                              
\multicolumn{1}{c}{}&[2-6] &9.5     & 4.02(2)   &27(3)     &4.3(2)     &-0.30(1)\\
\multicolumn{1}{c}{}&[2-7] &299.1   &4.154(2)   &5.44(2)   &0          &-0.391(3)\\
\multicolumn{1}{c}{}&[2-7] &31.3    &4.122(8)   &17(1)     &3.5(1)     &-0.336(4)\\
\multicolumn{1}{c}{}&[2-8] &451.6   &4.133(1)   &5.61(2)   &0          &-0.39(2)\\
\multicolumn{1}{c}{}&[2-8] &45.7    &4.14(3)    &14.7(5)   &3.1(1)     &-0.347(2)\\
\multicolumn{1}{c}{}&[2-9] &742.3   &4.113(1)   &5.82(2)   &0          &-0.3866(2)\\
\multicolumn{1}{c}{}&[2-9] &46.9    &4.147(2)   &14.3(4)   &3.09(8)    &-0.350(1)\\                                              
\multicolumn{1}{c}{}&[2-10]&1475.1  &4.090(1)   &6.09(2)   &0          &-0.3836(2)\\                                               
\multicolumn{1}{c}{}&[2-10]&77.6    &4.140(1)   &15.7(3)   &3.34(6)    &-0.343(1)\\                                                
\multicolumn{1}{c}{}&[3,9] &27.8    &4.106(1)   &7.9(1)    &0          &-0.374(1)\\                                     
\multicolumn{1}{c}{}&[3-9] &7.5     &4.128(4)   &10.6(6)   &1.5(3)     &-0.363(3)\\                                               
\multicolumn{1}{c}{}&[4-9] &13.7    &4.110(2)   &8.8(2)    &0          &-0.37(1)\\  
\multicolumn{1}{c}{}&[4-9] &4.5     &4.14(1)    &13(1)     &3.1(9)     &-0.356(5)\\
%\multicolumn{1}{c}{} &[3-10]  &208.4&0.041&&0.086& 0.0,-0.369\\ 
%\multicolumn{1}{c}{} &[3-10] &54.5&0.041&0.163&2.97&-0.342\\
\multicolumn{1}{c}{} &[4-10] &119.8  &4.103(2)  &10.4(2)   &0          &-0.362(1)\\            
\multicolumn{1}{c}{} &[4-10] &37.2   &4.159(4)  &19(1)     &5.9(5)     &-0.331(4)\\
\multicolumn{1}{c}{} &[5-10] &65.7   &4.1(2)    &13(1)     &0          &-0.350(3)\\                                                
\multicolumn{1}{c}{} &[5-10] &10.4   &4.190(3)  &35(4)     &15(2)      &-0.29(1)\\                                                 
% &50.6   & 3.7(15)&3.0(2.7) &0.53(36)  &-0.353(1)\\
%\multicolumn{1}{c}{}&[5-11] &197.9&0.041&0.178&0&-0.334\\
%\multicolumn{1}{c}{}&[5-11] &23.4&0.042&0.536&7.25&-0.257\\
%\multicolumn{1}{c}{} &[6-11]&63.9   &4.2     &30(1)    & 0        &-0.303(4)\\
%\multicolumn{1}{c}{} &[6-11]&59.4   &3.5(24) &2.9(32)  & 0.71(39) &-0.328(4)\\
			\end{tabular}		
		\end{ruledtabular}
 \label{Table1}
\vspace{-1.5 cm} 
%  \end{minipage}
%  \end{flushleft} 
\end{wraptable}
%At fixed temperature $T$, the Polyakov loop correlators address the free energy of a system of two static color charges coupled to a heatbath.
The two point Polyakov-loop correlator is the partition function of the string within the transfer matrix interpertation~\cite{Luscher:2002qv} 
%---------------------------------------------
\begin{align}
\mathcal{P}_{\rm{2Q}} =& \int d[U] \,P(0)\,P^{\dagger}(R)\, \mathrm{exp}(-S_{w}),  \notag\\
=& \quad\mathrm{exp}(-V(R,T)/T),
\label{COR}  
\end{align}
%---------------------------------------------  
%---------------------------------------------
where the Polyakov loop on the lattice is defined as 
%---------------------------------------------
\begin{equation}
  P(\vec{r}_{i}) = \frac{1}{3}\mbox{Tr} \left[ \prod^{N_{t}}_{n_{t=1}}U_{\mu=4}(\vec{r}_{i},n_{t}) \right],
\end{equation}
%---------------------------------------------
  The Monte-Carlo evaluation of the $Q\bar{Q}$ static potential at each $R$ is calculated through the correlator Eq.\eqref{COR} after averaging the time links ~\cite{Ph}. The $Q\bar{Q}$ potential data are fitted to the static potential of rigid string modified by the Gaussian curvature Eq.~\eqref{20}; However, to succinctly disclose the influence of the novel topological term we consider the action augmented by the next-to-leading contribution of NG action, separately.The Monte-Carlo evaluation of the $Q\bar{Q}$ static potential at each $R$ is calculated through the correlator Eq.\eqref{COR} after averaging the time links ~\cite{Ph}. The $Q\bar{Q}$ potential data are fitted to the static potential of rigid string modified by the Gaussian curvature Eq.~\eqref{20}; However, to succinctly disclose the features of the novel topological term we consider the action augmented by the next-to-leading contribution of NG action, separately.

  The lattice employed in this investigation is of a typical spatial size of $3.6^3 \times 8$ $\rm{fm^4}$ with a lattice spacing $a=0.1$ fm corresponding to coupling $\beta = 6.00$ and temperatures $T/T_c =0.9$. The gauge configurations were generated using the standard Wilson gauge-action employing a pseudo-heatbath algorithm~\cite{FHKP} updating to the corresponding three $SU(2)$ subgroup elements~\cite{Marinari}. Each update step/sweep consists of one heatbath and 5 micro-canonical reflections. The measurements are taken after 2000 thermalization sweeps on 500 bins. Each bin consists of 20 measurements separated by 70 sweeps of updates.
  
  Table.~\ref{Table1} summarizes values of $\chi^2$ and the fit parameters of the potential model of the static meson Eq.~\eqref{20}. Data bins consist of string segments of different length. For some selected fit intervals the topological term is switched off keeping $\theta=0$  to gain insight comparing the purely smooth string.
  
\begin{figure}
%    \vspace{- cm}
    \begin{flushleft}
%\vspace{-0.1 cm}      
\subfigure[]{\includegraphics[scale=0.72]{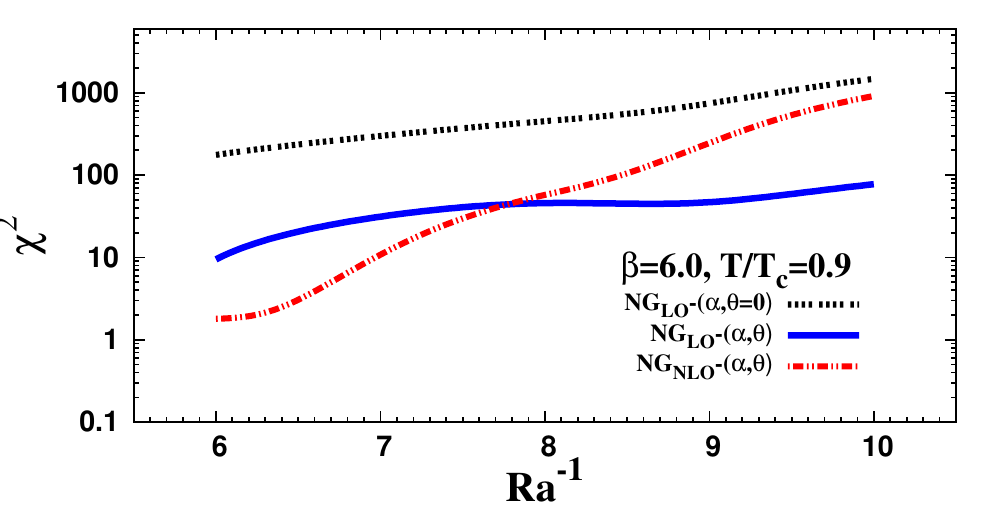}}\\
\subfigure[]{\includegraphics[scale=0.72]{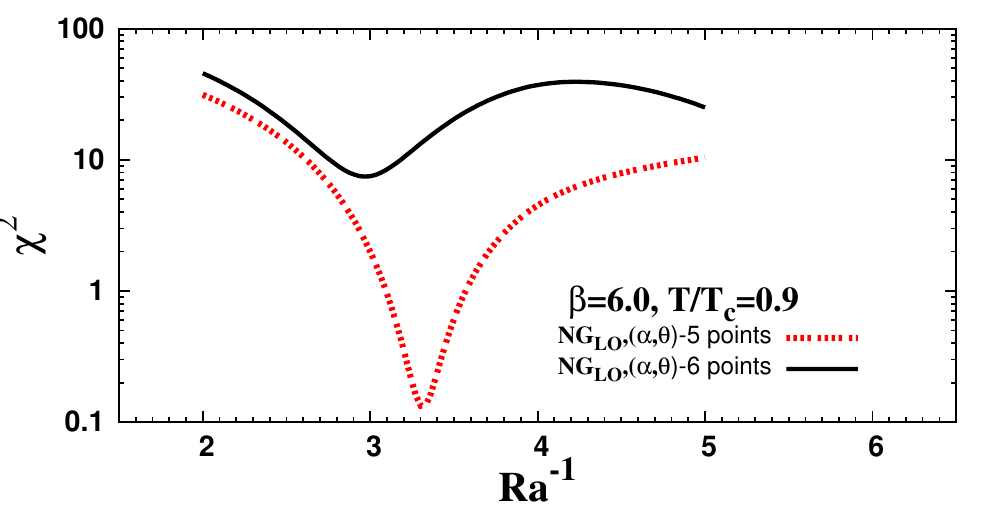}}
%\subfigure[]{\includegraphics[scale=0.72]{theta.eps}}
\end{flushleft}
\vspace{-9.7 cm}
\hspace{7.5 cm}
%\begin{flushright}
  \subfigure[]{\includegraphics[scale=0.77]{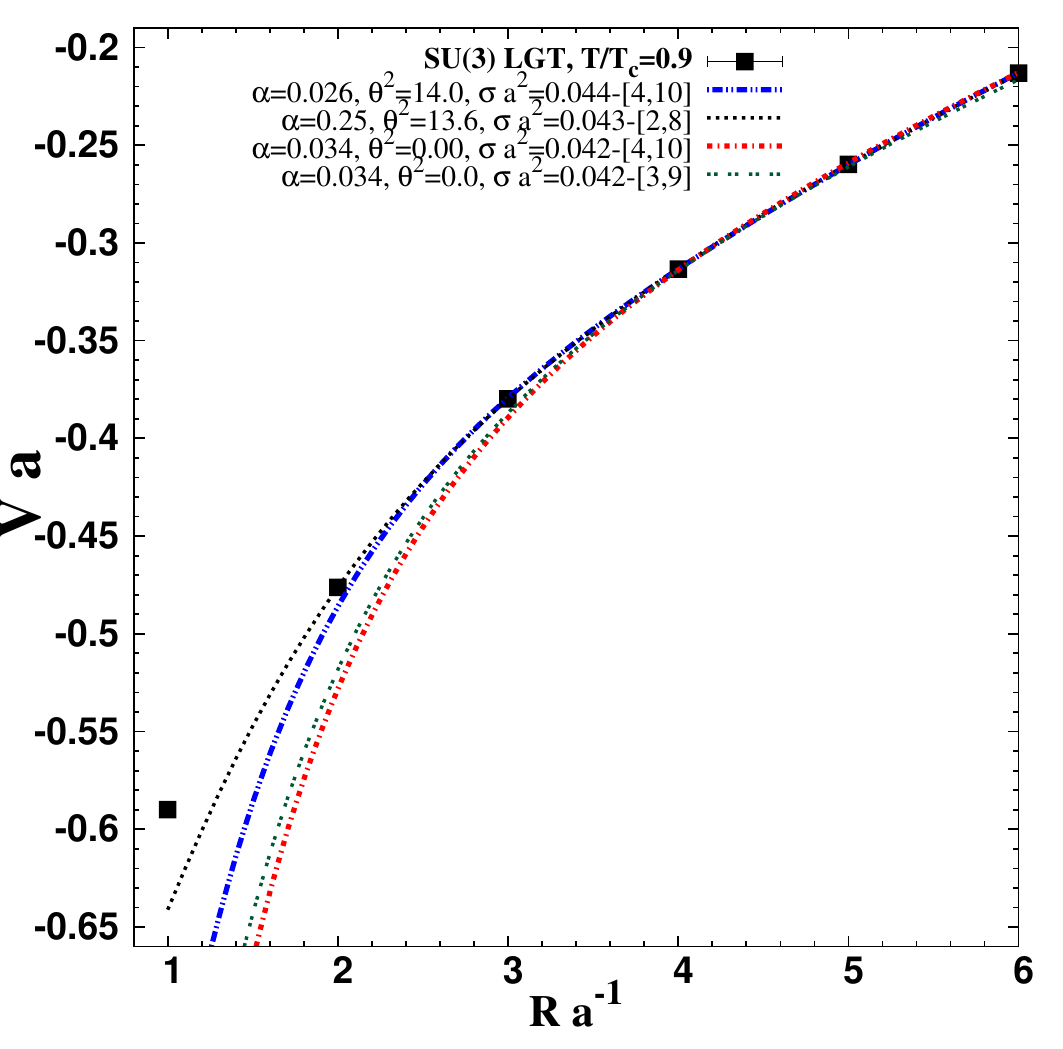}}
%\end{flushright}  
%  \includegraphics[scale=0.9]{Pot_Alpha_beta1.eps} 
  \caption{(a)Compares $\chi^{2}$ over fit regions $R\in[2,L_{\rm{M}}]$, the models describe the smooth string Eq.~\eqref{20} and Eq.~\eqref{POTNGNLO} with Gaussian-curvature $\theta^2 \ne 0 $ and $\theta^2=0$. (b) Same as in (a); however, the fit intervals are binned into segments of 5 and 6 points. (c)The lines display the fit of the lattice data of the static $Q\bar{Q}$ potential to either of smooth string models Eq.~\eqref{20} or Eq.~\eqref{POTNGNLO}.}  
\label{Fig2}
\end{figure}

  Figure~\ref{Fig2}-(a) confront the $\chi^{2}$ returned from fits of Eq.~\eqref{20} to $Q\bar{Q}$ potential data over fit intervals $R \in [2,L_{max}]$. A plummet in the values of residuals shows up with the involvement of the topological term $\theta\ne 0$ towards a better match with the lattice d. Moreover, the (DOF) $\chi^{2}_{\rm{dof}}$ appears to level-off with more points being contained in the fit interval (up to $L_{M}=10$). The pure smooth string model poorly fit the data as the very high values of $\chi^{2}$ expose. However, the influence of the Gauss-Bonnet term is substantial and overtakes the short distance scales, as depicted also in Fig.~\ref{Fig2}-(b), around $R=0.3$ fm and $R=0.4$ fm.
  
   The contribution to the string potential owing to the four-derivative term Eq.~\eqref{LW}, the couplings $\kappa_1, \kappa_2$ coincide with that of the Nambu-Goto (NG) action, is calculated using a free string propagator and $\zeta$ regularization~\cite{PhysRevD.27.2944,Billo:2012da}
%  compares the fits of data bins consisting 5 and 6 points with the x-axis co-ordinate presenting the smallest $Q\bar{Q}$ separation.  The best obtained fits for string segments of length corresponding to 5 and 6 points occur and evidently shifts the static potential curve 
\vspace{-0.03 cm}
\begin{equation}
\begin{split}
&  V(R,T;\alpha,\gamma,\mu)=V^{R}+V_{n\ell o}, \label{POTNGNLO}\\
&V_{n\ell o}=-T \log  \left( 1-\dfrac{(d-2)\pi^{2}T}{1152 \sigma_{o}R^{3}}  \left[2 E_4(\tau)+(d-4)E_{2}^{2}(\tau)\right] \right),%\label{NLO}
%+b_2 (d-2) \dfrac{\pi^{3} L_T}{60 R^{4}}E_{4}\left(\tau\right)+\dfrac{-b_4 (d-2)\pi^{5}L_{T}}{126  R^{6}} E_{6}\left(\tau\right),
\end{split}
\end{equation} 
\vspace{-0.8 cm}
 \begin{wraptable}{r}{9.0 cm}
    \begin{centering}
  \caption{Enlisted are the returned values of the fit parameters of the model Eq.~\eqref{POTNGNLO} to the static $Q\bar{Q}$ data.}      
		\begin{ruledtabular}
			\begin{tabular}{cc|cccccccccc}
%			  \multirow{2}{*}{\begin{turn}{90}\hspace{1cm}$ V_{{n\ell o}}^{R,b_2} $ \end{turn}} &\tiny{Fit Interval}
				&\multicolumn{2}{c}{\tiny{Fit Parameters, $T/T_{c}=0.9$}} &\multicolumn{2}{c}{}\\
				%&\multirow{2}{*}{$\chi^{2}$} &\multirow{2}{*} {$\mu$}
				%&\multirow{2}{*}{$\chi^{2}$} &\multirow{2}{*} {$\mu$}\\
				&$R \in I$
				&\multicolumn{1}{c}{$\chi^{2} $} &\multicolumn{1}{c}{$\sigma_{0} (fm^{-2})$ } 
			        &\multicolumn{1}{c}{$\alpha/a^2$}&\multicolumn{1}{c}{$\theta^2$} &\multicolumn{1}{c}{$\mu$}\\
				\hline
%\multirow{10}{*}{\begin{turn}{90} $V_{LO}$ \end{turn}}				
%\multirow{10}{*}{\begin{turn}{90}\hspace{1cm}$V_{{n\ell o}}+V^{R}+V^{b_2}$ \end{turn}}
%\multirow{10}{*}{\begin{turn}{90}\hspace{1cm}$V_{{n\ell o}}+V^{R}+V^{b_2}$ \end{turn}}				
% &0.045 &1083.06 &-0.446299\\ 				
%\multicolumn{1}{c}{} &0.038 & 1099.03   &      4.73    &   96.54      &  58.34        \\
\multicolumn{1}{c}{} &\tiny{$[L_{m},L_{M}]$}  &    &   \\			
\multicolumn{1}{c}{} &[2,6] &$ 4.7 \times 10^{5}$ & 2.87(3)& 0.03(1)  &0 &-0.3180(1)\\
\multicolumn{1}{c}{} &[2,6]&  1.8  & 4.28(1)& 1.6(1)& 22.4(5)& -0.4017(3)\\
%                                     0.00008890  0.0006733  0.543733 0.00028987},
                                      %{0.0138042, 148.163, 2.03798, 2.887}
\multicolumn{1}{c}{} &[2,7]& 10.8 & 4.260(5)& 1.49(5)& 23.4(5)& -0.4010(2)\\
%                           {0.0000541758 0.0005 499  0.50331, 0.0002 184026}
\multicolumn{1}{c}{} &[2,8]& 57.5&4.235(4)&1.34(4)& 25.2(5)& -0.4001(1)\\
%                            0.000404376, 0.545397, 0.000142663}   
\multicolumn{1}{c}{} &[2,9] &245 &4.204(3)&1.12(3)&28.6&-0.3989(1)\\
%                           0.000028131, 0.000331068,0.697068,      0.000120615},
\multicolumn{1}{c}{} &[2,10] &914 &4.165(2)   &0.84(3)  &36(1)&- 0.3971(1)\\
%                           0.0000205665, 0.000274648,1.1824, 0.000109224 
\multicolumn{1}{c}{} &[3,9] &1.0 $\times 10^{4}$   & 4.0(4) & 0.24(+39)& 0&  -0.39(1)\\
                            %{0.0039492        3.95304, 0., 0.0118664}
\multicolumn{1}{c}{} &[3,9]&6.0        & 4.432(2)& 20(1) & 13.2(1)& -0.334(4)\\                                         
                            %&19.6&    0.0445 & 0.31  & 6.3 &-0.306\\
%\multicolumn{1}{c}{} &[3,10]&$1.1\times 10^4$ & 4.0(4)& 0.2($+30.7$) &0 &-0.39(1)\\
%                           {0.37613 3.78328, 0.,      0.0113235}
\multicolumn{1}{c}{} &[3,10]&32      & 4.425(1) & 25(1)& 13.6(1)& -0.320(3)\\
%\multicolumn{1}{c}{} &[4,9] &160     & 4.16   & 1 $\pm$ 24.8   &0    &-0.4(1)\\
                            %{0.0417876, 24.8862, 0.,             0.132862}
\multicolumn{1}{c}{} &[4,9] &3.1  & 4.424(6)& 17(2)& 12.2(6) & -0.343(6)\\
\multicolumn{1}{c}{} &[4,10]&487  & 4.214(5) & 3.4(2) & 0& -0.3880(8)\\                      
\multicolumn{1}{c}{} &[4,10]&30.6 &      4.427(3)& 26(2) & 14.0(4)& -0.316(5)\\                                   
\multicolumn{1}{c}{} &[5-10]&119.6&      4.34 (3)  & 11 (5)    & 0& -0.359(2)\\
%                                 {0.0000386339, 0.0047469, 0.,           0.00181862}
\multicolumn{1}{c}{} &[5-10]&9.9  &      4.44(2)& 40($\pm 30$)& 21(1)& -0.28(2)\\
%\multicolumn{1}{c}{} &[6,11]& 79.3 &  0.0440 (4)& 0.31 (4)& 0  & -0.300(2)\\
%                                   {0.0000402075, 0.00496937, 0., 0.00191028}
%\multicolumn{1}{c}{} &[6,11]& {81.9173, {\[Sigma] -> 0.0440229, \[Alpha] -> 0.304135, \[Beta] -> 0.400002, R0 -> -0.302278}}
%                             5.6  &  0.0443 & 1.00 & 10.5 & -0.203\\                                   
			\end{tabular}		
		\end{ruledtabular}                              
 \label{Table2}
\end{centering} 
\end{wraptable}
 \vspace{0.0 cm}
 
 where $E_{2n}(\tau)$ is Eisenstein series \cite{Eisen} and $\tau=\dfrac{2L_{T}}{R}$ is the modular parameter of the cylinderical-sheet.
 
  The returned fit parameters and the residuals considering Eq.~\eqref{POTNGNLO} are laid out in Table.~\ref{Table2}. The comparison between fits in Table.~\ref{Table1} debuts two distinct effects of including the NG self-interactions term. As depicted in Fig.~\ref{Fig2}-(b) a further decrease in $\chi^{2}$ from fits over $R \in [0.2,L_{M}]$ fm occurs up to separation distances $L_{M}=0.8$ fm. The rise in $\chi^2$ at $R \ge 0.8$ may suggest the need to include more boundary terms in fit scheme~\cite{Bakry:2020flt}. Apart from this, the fit drastically escalates while the coupling is switched off $\theta^2=0$.

  In Fig.~\ref{Fig2}-(c) the fitted static $Q\bar{Q}$ potential curves  Eq.~\eqref{POTNGNLO} are presented over source separation interval $R\in[0.4,1]$ fm. Interestingly, the curves corresponding to the model coupled to the Gaussian curvature spontaneously tend towards a closer match with the data points at shorter string length $R \le 0.4$ fm. The model's fits over intervals $R \in [0.2,0.8]$ fm and $R \in [0.3,0.9]$ fm illustrate the considerable influence of the topological term at short length scales. %The comparison reveals that smooth string model coupled to Gaussian curvature compares favorably to Monte-Carlo data for source separation $R \le 0.5$ fm.
 \vspace{-0.2 cm}
\begin{wrapfigure}{r}{0.487\textwidth}
%\begin{flushright}  
  \includegraphics[scale=0.93]{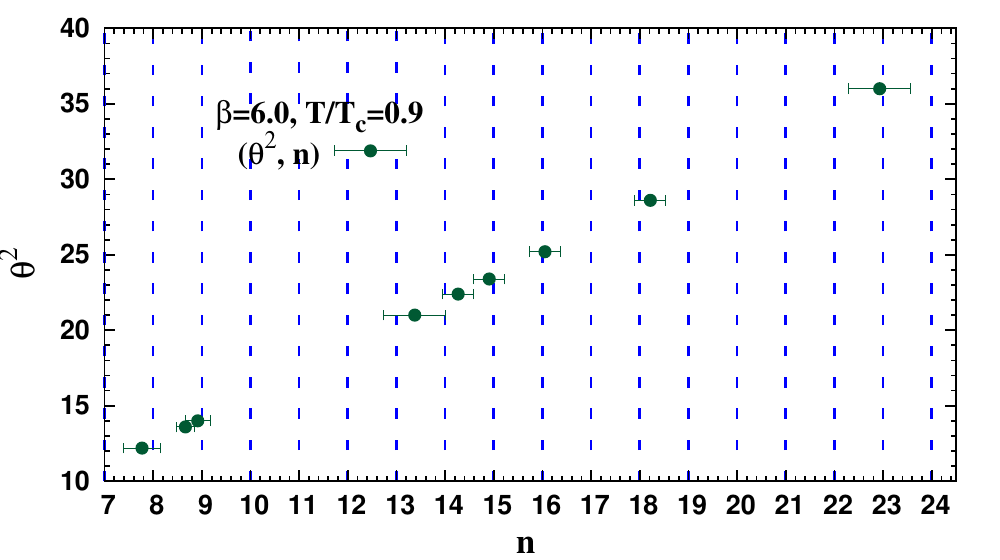}
\vspace{-0.8 cm}  
\caption{\label{Fig3}The worldsheet's winding mode $n=\dfrac{2\theta^2}{\pi}$ measured using the fit to Eq.~\eqref{POTNGNLO} in Table~\ref{Table2}.}
%  -fit intervals in Tables~\ref{Table2}.
%\end{flushright}
%\vspace{-1.0 cm}  
\end{wrapfigure}
\vspace{-0.9 cm}

~~Almost all the retrieved values of the coupling $2 \theta^2$ turn out to be integer multiples of $n \pi$. The remarkable quantization could be interpreted as emerging from the winding number of the surface~\cite{Lanzat, Dub}. The total surface integral of the curvature being repeated $n$ winding times~\cite{Lanzat}. This is supported by (sign-independent) quadratic potential in the coupling and may have to do with the proclaimed quasi-particle modes on the worldsheet~\cite{Dub}. In Table.~\ref{Table1} the returned quantum numbers are at least $n=1,2$ otherwise $n=5$ at $R \ge 0.5$.

   With the string's self-interactions been encompassed in the model Eq.~\eqref{POTNGNLO}, the plot in Fig.~\ref{Fig3} affirms the same observation. However, the corresponding values in Table~\ref{Table2} indicate higher values, $n=8$ to $n=23$. This seems compatible with a worldsheet self-intersecting $n$ times induced by the self-interaction term. 
%   The returned values of the coupling $\theta^2$ seem to roughly equal integer or half integer multiples of $\pi$. 

  The manifest values of $\sigma_{0}$ in Table~\ref{Table1} do not match $T=0$ string tension $\sigma_{0}a^{2}=0.0445(5)$~\cite{Bakry:2020flt}. The coupling to the Gaussian-curvature, nevertheless, reveals no significant effects on the value of string tension. The fits in Table~\ref{Table2} points out to the correct value of $\sigma_{0}a^2$ and the corresponding dependency on the temperature~\cite{Bakry:2020flt} while the smooth string action being augmented with the two-loop terms of the NG action.
\paragraph{{\bf \large{Conclusion}}}
The static potential model of open smooth strings coupled to the geometric-invariant of the Gaussian curvature have been discussed within the non-perturbative region of hadronic forces. In the vicinity of critical point, $T/T_{c}=0.9$, fits to $Q\bar{Q}$ potential data in pure $SU(3)$ YM-theory including the Gaussian-curvature model sustains a significant improvement in the match to Monte-Carlo data over short and intermediate color source separation distances $R \in [0.2, 1.0]$. The coupling parameter of topological term debuts a quantisation number that could be interpreted as the winding modes or self-intersection/winding times of the string's worldsheets.   
\paragraph{{\bf Acknowledgment}}
\footnotesize{This work has been funded by the Chinese Academy of Sciences President’s International Fellowship Initiative grants No.2015PM062 and No.2016PM043, the Recruitment Program of  Foreign Experts, the Polish National Science Centre (NCN) grant 2016/23/B/ST2/00692, NSFC grants (Nos. 11035006, 11175215, 11175220) and the Hundred Talent Program of the Chinese Academy of Sciences(Y101020BR0).}
\vspace{-0.7 cm}
%%%%%%%%%%%%%%%%%%%%%%%%%%%%%%%%%%%%%%%%%%%%%%%%%%%%%%%%%%%%%%%%%%%%%%%%%%%%

\end{document}